\documentclass[a4paper]{article}
\usepackage{setspace}
\usepackage{amsmath}
\usepackage{graphicx,subfigure}
\usepackage{epstopdf}
\usepackage{upgreek}
\usepackage[margin=2.0 cm]{geometry}

\begin{document}
\begin{center}

\large \textbf{Defect-polymorphism controlled electrophoretic propulsion of anisometric microparticles in a nematic liquid crystal}

\hfill \break
Devika V S, Dinesh Kumar Sahu, Ravi Kumar Pujala and Surajit Dhara$^{*}$

\hfill \break

\end{center}

\begin{abstract}
 
 Nontrivial shape of colloidal particles create complex elastic distortions and topological defects in liquid crystals and play a key role in governing their electrophoretic propulsion through the medium. 
   Here, we report experimental results on defects and electrophoretic transport of anisometric (snowman-shaped) dielectric particles subjected to an alternating electric field perpendicular to the director in  a nematic liquid crystal. We demonstrate that the shape asymmetry gives rise to defect-polymorphism by nucleating point or ring defects at multiple locations on the particle and controls the direction as well as the magnitude of the electrophoretic propulsion.
 Our findings unveil a novel degree of freedom in translocating microparticles in liquid crystals for applications in microfluidics, controlled transport and assembly. \\\\\\\\\\\\\\\\\\\\
\end{abstract}
  
\noindent\rule{16cm}{0.5pt}\\
Devika V S, Dinesh Kumar Sahu and Surajit Dhara \\ 
  School of Physics, University of Hyderabad, Hyderabad-500046, India \\ 
 Dr. Ravi Kumar Pujala \\
Soft and Active Matter Group, Department of Physics, Indian Institute of Science Education and Research (IISER), Tirupati, Andhra Pradesh-517507, India  \\
 $^{*}$Corresponding author:surajit@uohyd.ac.in \\

\section{Introduction}
 Active control of microparticles and fluids at microscale by transducing the energy of an external electric field has been a subject of intense study owing to its wide applications, ranging from directed assembly of colloids to sorting of macromolecules, biomolecules and microfluidics~\cite{morgan,ramos, drop,alex,div2,rc}. In an isotropic suspension (e.g., aqueous) particles are transported due to the classical (linear) or the induced-charge (nonlinear) electrophoresis~\cite{mur,tm,baz1,baz2,baz3}. In liquid crystals (LCs), the mechanism is markedly different from their isotropic counterparts as the particles distort the medium, and break the symmetry of near-field (local) director by nucleating topological defects~\cite{lub,stark}. 
  A spherical particle with homeotropic surface anchoring in a nematic liquid crystal (NLC) nucleates either a point defect known as hyperbolic hedgehog (strength, $m=-1$) or a disclination ring defect ($m=-1/2$) called Saturn ring~\cite{igor,im,is}. The particles, accompanying point or ring defects are known as elastic dipoles or quadrupoles, respectively, in analogy with electrostatics~\cite{lub,stark}. 
  
     The applied electric field in NLCs creates a spatial charge separation across the particle in distorted regions with a charge density $\rho$  proportional to the strength of the electric field ${\bf E}$. In response, an electrostatic force, $f\propto\rho E\sim E^2$ starts acting on the induced space charge cloud around the particle. This force drives the LC molecules surrounding the particles, a phenomenon known as LC-enabled electro-osmosis (LCEO)~\cite{od,od1,oleg}.  For spherical dipolar particles, the fore-aft symmetry of the LCEO flow is broken due to the asymmetric director configuration; as a result the particles propel along the far field director $\bf{\hat{n}}$ (average direction of molecular orientation). Whereas, in case of quadrupolar particles, the fore-aft symmetry of LCEO flow is preserved and the particles do not propel~\cite{oleg1,oleg2}, except for metal-dielectric Janus particles reported recently~\cite{sd,sd1,sd2}. 
         
     While the transport of spherical particles in NLCs has been extensively investigated, the recent advances in colloidal synthesis as well as the advent of 3D printing have made it possible to fabricate microparticles with complex geometries~\cite{david1,david2,liu,3D}. It is known that the shape asymmetry of particles play an important role in inducing variety of defects, complex assemblies~\cite{science,jaya,sh-1,sh-2,sh-3,sh-4,sh-5,sh-6,rasi} as well as transport in NLCs~\cite{sag1,sag2,sag3,rasna}. However, controlling the  electrokinetics by modulating the defects around the particles in NLCs is as yet unexplored. In this paper, we report experimental studies on defects and electrokinetics of snowman-shaped particles in an NLC. We show that the morphology dependent defect-polymorphism leads to rich electrokinetics which are potential for applications in microfluidics, directed-assembly and transport.
      
  
   \section{Results and discussion}

\begin{figure*}[ht]
\centering
\includegraphics[scale=0.22]{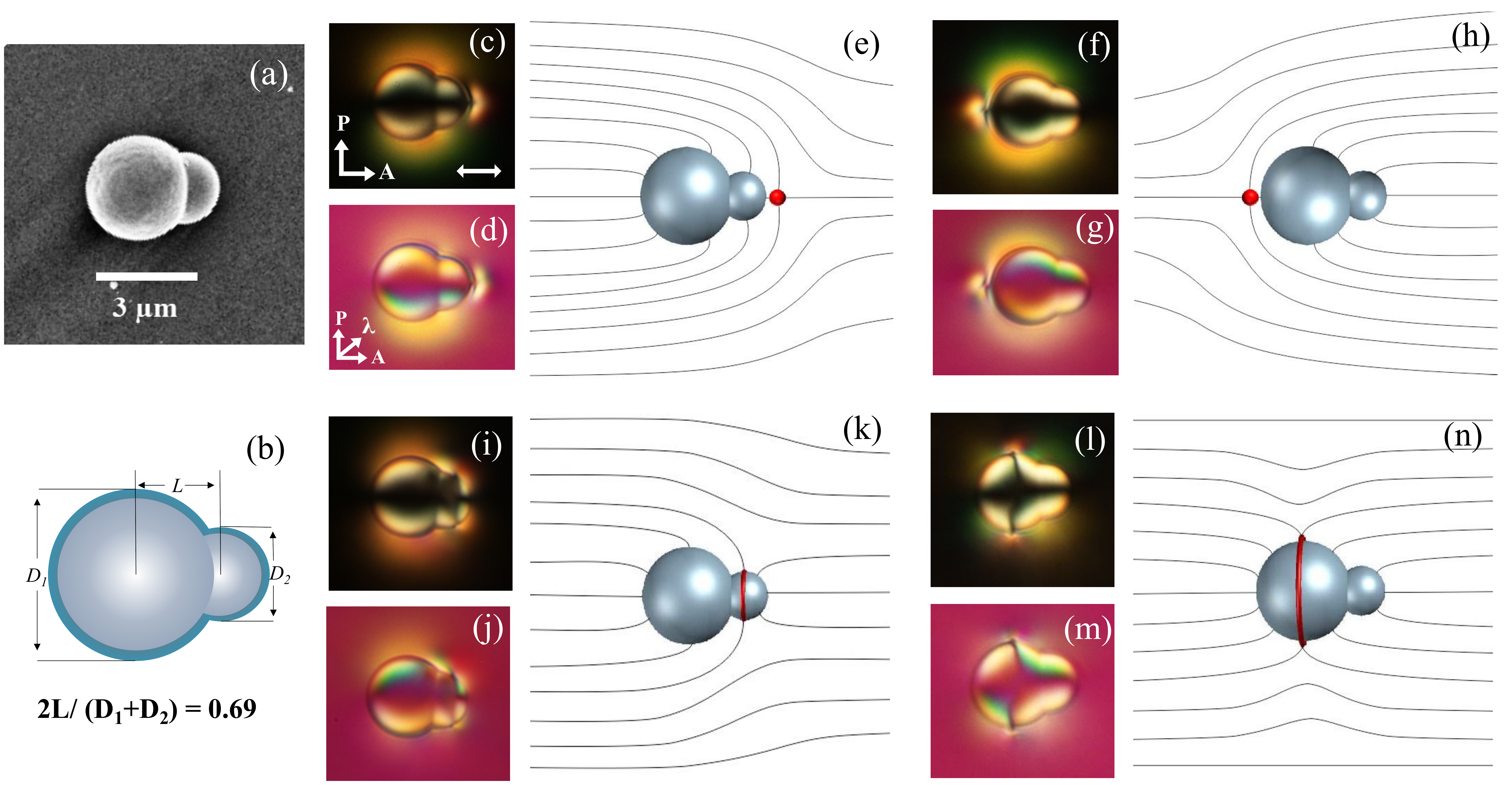}

\caption{(a) Scanning electron microscope (SEM) image of a SM particle. (b) Cross-section of the core-shell structure of a particle. Thickness of the silica shell (Cyan) is 50nm. Separation factor $2L/(D_1+D_2)=0.69$.  Polarising optical microscope (POM) images of SM particles in NLC with (c) point defect on the small lobe (f) point defect on the big lobe (i) Saturn ring defect on the small lobe, and a (l) Saturn ring defect on the big lobe. Figures (d,g,j,m) show corresponding images taken by inserting a $\lambda$-plate (530 nm) at an angle of 45$^{\circ}$ between the sample and the analyser. (e,h,k,n) Schematic director profiles around the particles. Red dot and red circles represent point and Saturn ring defects. Double headed arrow on represents the nematic director. P, A represents polariser and analyser.} 
\label{fig:figure1}
\end{figure*}

\begin{figure*}[ht]
\centering
\includegraphics[scale=0.19]{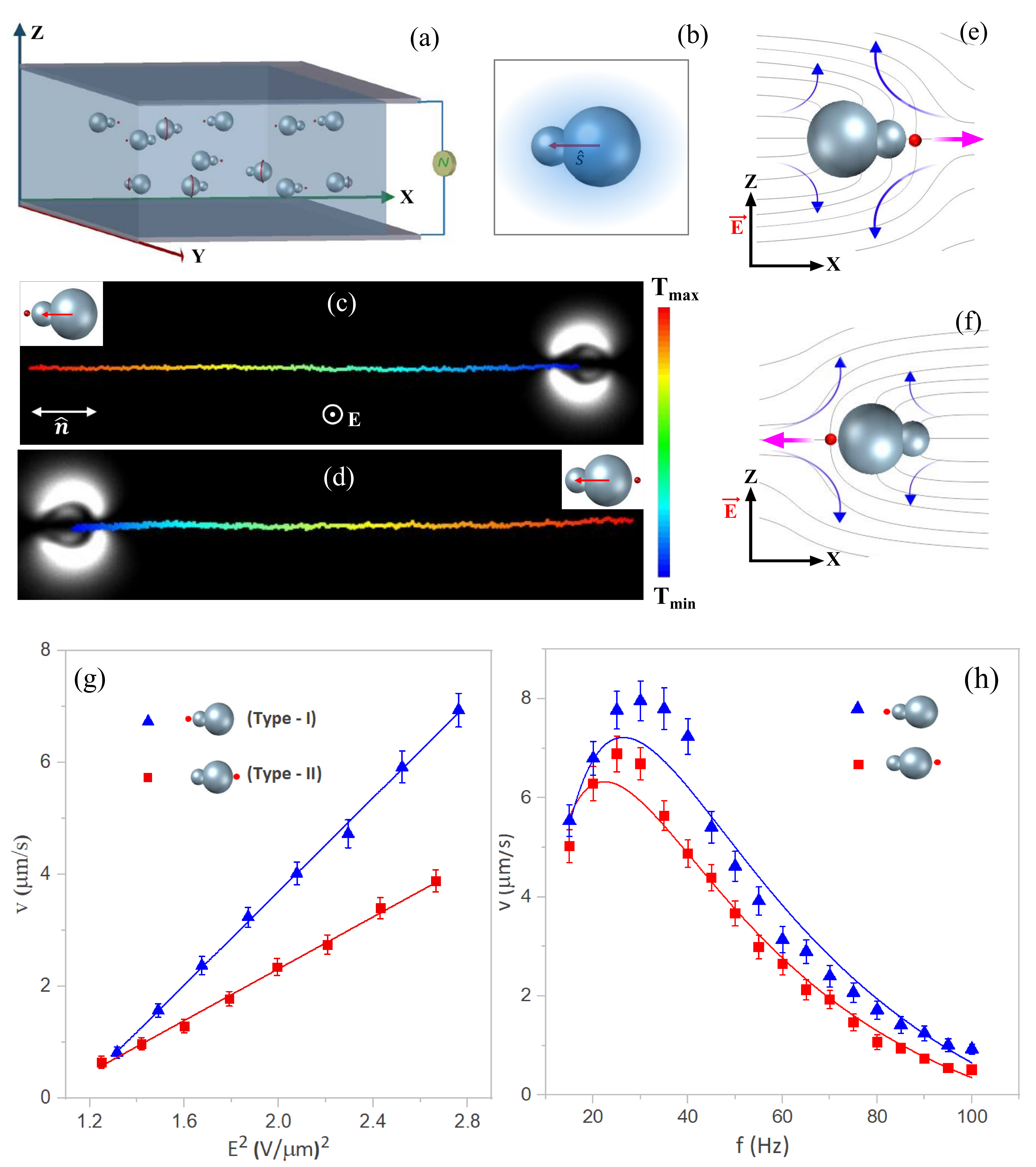}

\caption{(a) SM particles in a planar cell with an ac field along $z$-axis.  Director ${\bf\hat{n}}$ is parallel to $x$-axis. (b) Polar vector ${\bf\hat{s}}$. (c,d) Time coded trajectories of propulsion of type-I and type-II dipolar SM particles (see Movie S1 and S2). Applied field E =  1.3 V$/\upmu m$ at 50 Hz.  Double-headed arrow represents the director ${\bf\hat{n}}$. (e,f) LCEO flows (puller type) marked by blue curved arrows. Magenta coloured arrows indicate propulsion direction. (g) Field dependent velocity of type-I  and type-II dipolar SM particles. Blue (red) line represent least square fits to Eq.(1) for type-I (type-II) particles with slopes 4.18 and 2.3 (V$/\upmu$m)\textsuperscript{2}, respectively (h) Frequency dependence of velocity of type-I  and type-II dipolar SM particles. Solid curves represent least square fits to Eq.(2). T\textsubscript{max}= 20 s and T\textsubscript{min}= 0 s. Error bar shows the standard deviation of the mean value. }  
\label{fig:figure2}
\end{figure*}

Snowman-shaped particles coated with DMOAP mostly orient with their long axis parallel to the far-field director {$\hat{\textbf{n}}$}, enforced by the rubbing direction. They spontaneously induce dipolar or quadrupolar type director configurations similar to that of spherical particles. In addition, they show defect polymorphism in which the point and ring defects nucleate either on the big or small lobe of the snowman (SM) as shown in Fig.\ref{fig:figure1} (c,f,i,l). The near-field director distortions and resulting defects are elucidated by inserting a full-wave retardation plate or $\lambda$-plate between the crossed polarisers (Figs.\ref{fig:figure1}(d,g,j,m)).
Two types of dipolar configurations are identified based on the position of the point defects. For type-I dipoles, the point defect is nucleated on the small lobe (Fig.\ref{fig:figure1} (e)) whereas, for type-II dipoles the point defect is nucleated on the big lobe (Fig.\ref{fig:figure1} (h)). Similarly, two quadrupolar type defect structures are observed depending on whether the Saturn ring is present on the small (type-I) or big (type-II) lobe of the particle as shown schematically in Fig.\ref{fig:figure1} (k,n). 
   
   The evolution of these defects can be arranged in a progressive manner as shown schematically in Fig. \ref{fig:figure4}(a). Starting from the left, the point defect is first nucleated on the small lobe (Fig.\ref{fig:figure4}(a-i)), then it slightly opens up to form a Saturn ring encircling the small lobe (Fig.\ref{fig:figure4}(a-ii)),  which then migrates to form a larger ring encircling the big lobe (Fig.\ref{fig:figure4}(a-iiii)). Finally, the larger ring collapses on the left side of the big lobe and transforms to a point defect (Fig.\ref{fig:figure4}(a-iv)). Such polymorphism in defects is allowed as a ring-defect of strength, $m=-1/2$ is topologically equivalent to a hyperbolic hedgehog defect of strength, $m=-1$. In effect, the shape asymmetry of SM particles give rise to rich defect polymorphism in NLCs.

\begin{figure*}[ht]
\centering
\includegraphics[scale=0.25]{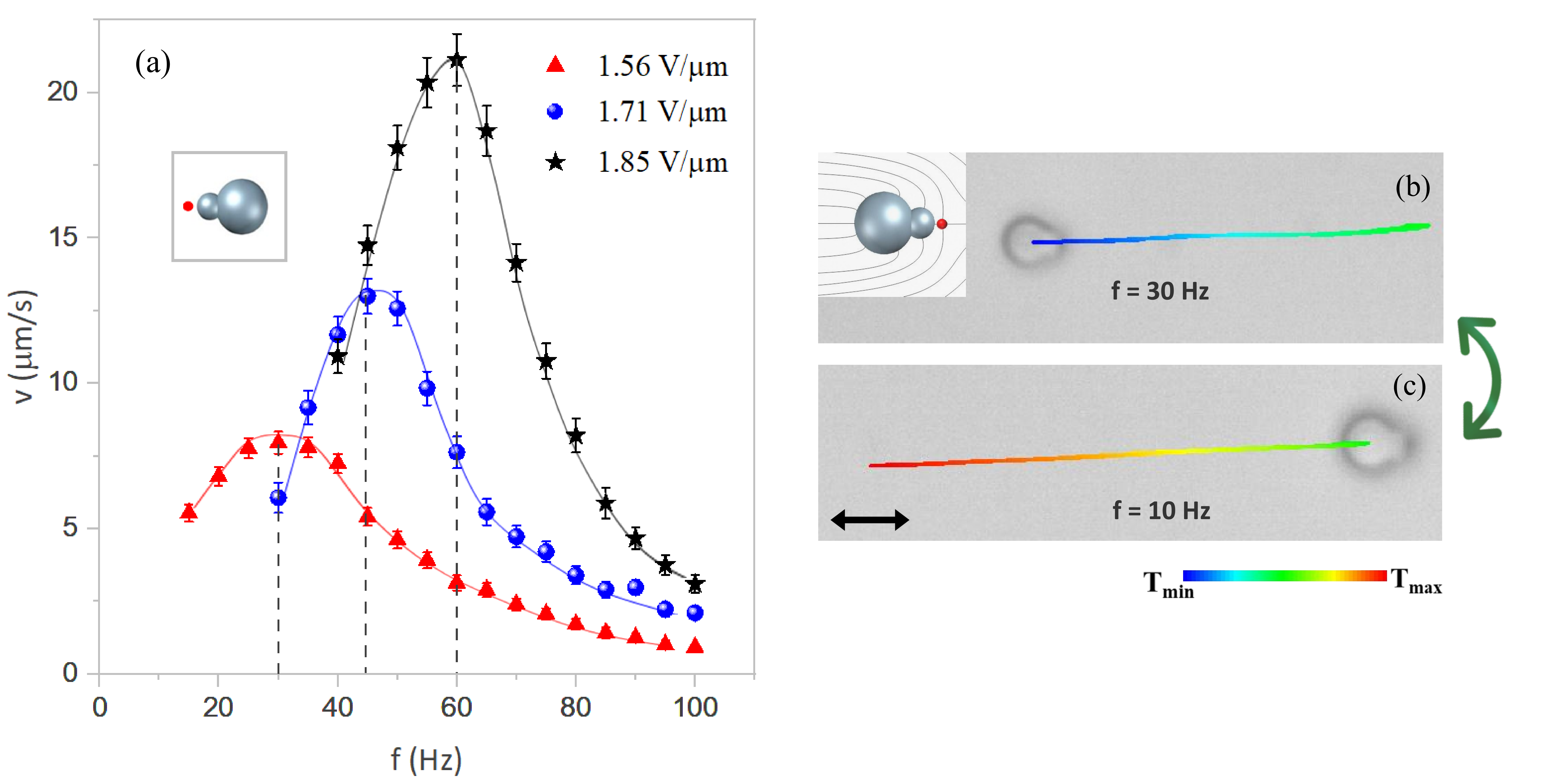}
\caption{(a) Frequency dependence of the electrophoretic velocity of SM dipolar particles at three different electric field values 1.56 V/$\upmu$m, 1.71 V/ $\upmu$m and 1.85 V/ $\upmu$m. Solid curves are guide to the eye. Vertical dotted lines indicate the peak frequency. (b,c) Reversal of direction of motion from ${\bf+\hat{s}}$ to ${\bf-\hat{s}}$ by changing the frequency from 30 Hz to 10 Hz, keeping the field constant at E = 1.3 V/$\upmu$m (see Movie S5). } 
\label{fig:figure3}
\end{figure*} 

 We study the electrokinetics of SM particles under ac (square wave) electric field \textbf{E}, applied perpendicular to the director as shown in Fig.\ref{fig:figure2}(a). The electric field does not affect the macroscopic director ${\bf\hat{n}}$ as the dielectric anisotropy $(\Delta \epsilon$) of the liquid crystal MJ-98468 is negative in the working frequency range (Fig. S2, Supporting Information). First, we consider type-I and type-II dipolar SM particles only.  Beyond a threshold electric field both the particles start propelling parallel to the far field director $\bf{\hat{n}}$, but in opposite directions as shown in Fig.\ref{fig:figure2}(c,d)) (see Movie S1,S2). Here, we define a vector ${\bf \hat{s}}$ (directed towards the centre of small lobe from the centre of big lobe) which indicates the shape polarity of the particle (Fig.\ref{fig:figure2}(b)). For type-I dipolar particles, the point defect on the small lobe leads the way, and the propelling direction is parallel to $\bf {\hat{s}}$ as shown in Fig.\ref{fig:figure2}(c) (particle moving from right to left). Whereas, for type-II dipolar particles, the point defect on the big lobe leads the motion and the propelling direction is opposite to $\bf{\hat{s}}$ as shown in Fig.\ref{fig:figure2}(d) (particle moving from left to right). The electrophoretic velocity of a spherical particle of radius $R$ is given by~\cite{oleg2}
 \begin{equation}
u=\alpha\frac{\epsilon_{0}\overline{\epsilon}R}{\eta}(\frac{\Delta\epsilon}{\overline{\epsilon}}-\frac{\Delta\sigma}{\overline{\sigma}})E^{2}
\end{equation}
where $\alpha$ is a constant, $\eta$ is the viscosity, $\overline{\epsilon} = (\epsilon_{\parallel} + \epsilon_{\perp})/2$ is the average dielectric permittivity and $\overline {\sigma}= (\sigma_{\parallel}+\sigma_{\perp})/2$ is the average conductivity of the LC medium. Usually the sign of the term ($\Delta\epsilon/\overline{\epsilon}-\Delta\sigma/\overline{\sigma}$) decides the direction of motion of the dipolar particles. If it is positive, the point defect leads the way and if it is negative the point defect trails behind the trajectory~\cite{satya}. For our liquid crystal (MJ98468) this quantity is positive in the working frequency range, hence the point defect leads the motion as expected (Fig.S2 (c,d)). An interesting feature of the transport of dipolar SM particles in NLC is that the direction of motion of the particles could be reversed without reversing the shape polarity vector ${\bf \hat{s}}$, in contrary to their response in aqueous suspensions~\cite{ning}. 
 
The field dependent velocities of type-I and type-II dipolar particles are notably different. Figure \ref{fig:figure2}(g) shows that the velocities of both types of dipolar SM particles are proportional to $E^2$, as expected in LC-enabled electrophoresis (LCEEP). But, the slope of type-I dipoles is almost double than that of type-II, which means that at a given field, type-I dipoles propel with twice the velocity of type-II dipoles. The mobility of dipolar spherical particles in NLCs is due to the breaking of the fore-aft symmetry of the surrounding LCEO flow owing to the asymmetric director structure.  For SM particles, the symmetry of the LCEO flow is broken due to the asymmetric director structure, and additionally by their shape asymmetry. Due to this combined effect, the type-I dipolar particles propel at higher velocities than type-II dipoles. In analogy with spherical particles~\cite{oleg2}, we schematically present LCEO flows around the dipolar SM particles in Fig.\ref{fig:figure2}(e,f). The flow is puller type with respect to the director for both types of dipoles, but the magnitude of the flow is stronger near the small lobe for type-I whereas it is stronger near the big lobe for type-II dipoles. The stronger flows near the respective lobes of the type-I and type-II dipoles control their direction of transport. The effect of shape asymmetry is also reflected in the frequency dependent velocity  as shown in Fig.\ref{fig:figure2}(h). The velocity of type-I dipolar particles is greater than that of the type-II dipoles at all frequencies and it can be approximately fitted to~\cite{baz1,oleg1} 
\begin{equation}
 v(\omega) = v_{0}\frac{\omega^{2}\tau_{e}^{2}}{(1+\omega^{2}\tau_{e}^{2})(1+\omega^{2}\tau_{p}^{2})}
 \end{equation}
where $\omega = 2 \pi f$,  $\tau_{e}$ and $\tau_{p}$ are the electrode and particle charging times that regulates the electro-osmotic flows surrounding the particles, and are related to the lower and upper limits of the driving frequencies. We obtained moderate fittings to Eq.(2) when compared to that of spherical particles~\cite{sd}. 

\begin{figure*}[ht]
\centering
\includegraphics[scale=0.22]{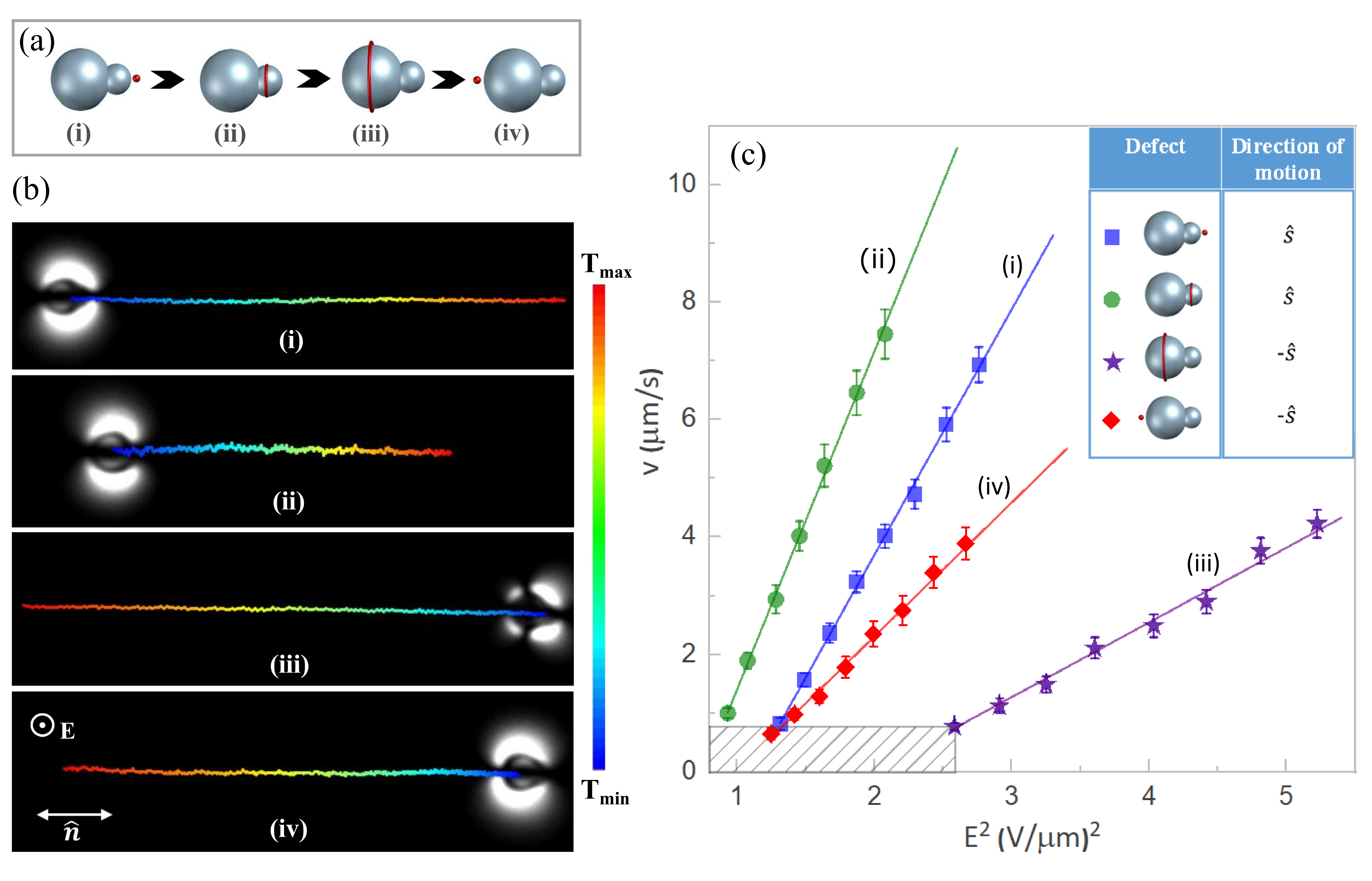}
\caption{(a) Arrangement of particles showing the progression of defects (left to right) from a point defect on the small lobe to a point defect on the big lobe through two successive ring-defects. (b) Time-coded trajectories of the particles as labelled from (i) to (vi). Note that particles with point  and ring  defects (Movies S1 and S2) on small lobe propel from left to right and particles with point and ring defects on big lobe ((Movies S3 and S4)) propel from the right to left. (e) Variation of velocities of four particles as a function of $E^{2}$ with slopes (i) 4.18 $\upmu$m\textsuperscript{3}V\textsuperscript{-2}s\textsuperscript{-1},  (ii) 5.76 $\upmu$m\textsuperscript{3}V\textsuperscript{-2}s\textsuperscript{-1}, (iii) 2.28 $\upmu$m\textsuperscript{3}V\textsuperscript{-2}s\textsuperscript{-1}, and (vi) 1.27 $\upmu$m\textsuperscript{3}V\textsuperscript{-2}s\textsuperscript{-1} as shown in diagram (a). Inset shows the direction of motion of respective particles either parallel or antiparallel to ${\bf\hat{s}}$. No electrophoretic transport is observed below the threshold fields (shaded region).} 
\label{fig:figure4}
\end{figure*}

 In what follows we  study the frequency dependent velocity of dipolar particles (type-I) at different electric fields. A comparison of frequency dependent velocity profiles at three fields, with a stepwise increase (0.15 V/$\upmu$m) from 1.56 to 1.85 V/$\upmu$m is shown in Fig. \ref{fig:figure3} (a).  
 With increasing field the frequency profile becomes sharper, and the peak frequency as well as the velocities rise to higher values. For example, the peak frequency is enhanced from 30 Hz to 60 Hz and the velocity (at the peak) increases from 7.5 $\upmu$m/s  to 22.2 $\upmu$m/s when the field is increased from 1.56 V/$\upmu$m to 1.85 V/$\upmu$m. The total shift in the peak frequency of SM particles ($\sim30$ Hz) is much greater than that was measured for spherical particles ($\sim10$ Hz) having similar size (Fig. S3, Supplementary Information). We also fit the velocity profiles at higher fields to Eq.(2) and observe that a moderate fitting to the data can be obtained when $\tau_{e}\approx\tau_{p}$ (Table-I, Supplementary Information). 
 It is analogous to the situation of frequency dependent electro-osmotic flows driven by ac fields at adjacent electrodes in microfluidic channels~\cite{baz2}. This suggests that at stronger fields the electro-osmotic flow between the electrodes dominates over the local flow surrounding the particles. 

  We further explored the low frequency regime ($< 20$Hz) of the velocity profile of SM dipoles. The frequency dependent velocity profile at \textbf{E}=1.56 V/$\upmu$m (Fig. \ref{fig:figure3}(a)) shows that the motion of the particle is stopped below 15 Hz. When the frequency is reduced nearly to 10 Hz, interestingly the SM particles start moving again but in the opposite direction of their earlier motion as shown in Fig. \ref{fig:figure3}(b) (see Movie S5). This unusual behavior of the particles at low frequencies points to an adept mechanism for direction reversal, which could otherwise be obtained only by changing the sign of the physical quantity $(\Delta\epsilon/\overline{\epsilon}-\Delta\sigma/\overline{\sigma})$~\cite{satya}. We confirm from an independent experiment that both the dielectric and conductivity anisotropies do not change sign when the frequency is changed from 20 Hz to 10 Hz (Fig. S2(c,d), Supporting Information). In the low frequency regime, particles can be transported due to the Carr-Helfrich instability~\cite{sasaki1,sasaki2,taka} in which the nematic forms thin cylindrical vortices, perturbing the initial director orientation~\cite{pg,blinov}. A careful observation suggests no visible electroconvection (so called Williams domains) occurred in the LC cell during the reverse motion of the particles. This reversal in motion, however, is limited to a very narrow frequency range (10-7 Hz), below which the particle's movement becomes random due to visible electrohydrodynamic instabilities. This means that the direction reversal occurs just before the onset of the Carr-Helfrich instability. The same phenomenon was observed at other fields, but at  different frequencies.

At the onset of the instability, the external electric field tends to align the director along the $z$-direction whereas, the elastic torque tries to restore its initial orientation. As a result of these two competing torques, slight director distortion starts developing across the sample. This director re-orientation in the vicinity of the hedgehog, which is still intact, induces a current flow \textbf{J} along the director ($x$-direction), leading to charge accumulation~\cite{acharya}. In response, an additional local field {\bf E}(x) develops along the $x$ direction. Migration of charges under {\bf E}(x) creates a local flow, which we anticipate, is the reason that causes the particles to propel in the opposite direction.  




Finally, we study the electrokinetics of type-I (Saturn ring on small lobe) and type-II (Saturn ring on big lobe) quadrupolar SM particles. The direction of the electric field is same as shown in Fig.\ref{fig:figure2}(a). 
   In sharp contrast to spherical quadrupoles which are immobile, SM quadrupolar particles propels parallel to the director ${\bf\hat{n}}$. Figures \ref{fig:figure4}(b)(ii,iii) present the time-coded trajectories of type-I and type-II quadrupolar SM particles. We also include the results of type-I and type-II  dipolar SM particles for a comparative discussion and greater insight. 
   SM particles with point or ring defects on the small lobe propel along the same direction (left to right) with the direction of motion parallel to  ${\bf +\hat{s}}$ (Fig.\ref{fig:figure4}(b)(i,ii)). This is because the ring defect on the small lobe creates elastic distortions somewhat similar to dipolar particles as evident from the two-lobe pattern observed under cross polarisers. 
    SM particles with point or ring defects on the big lobe propel with the direction of motion from right to left  i.e., the direction of motion is antiparallel to ${\bf\hat{s}}$ (Fig.\ref{fig:figure4}(b)(iii,iv)). It should be noted that in contrast to spherical particles, the four lobe pattern of type-II SM quadrupoles is asymmetric (Fig.\ref{fig:figure4}(b)(iii)). As mentioned earlier, spherical quadrupolar particles do not propel due to the fore-aft symmetry of the LCEO flow. In our case, the quadrupolar symmetry of the director field (Fig.\ref{fig:figure1}(k,n)) and hence, the fore-aft symmetry of the LCEO flow is broken due to the inherent shape-asymmetry of the particles. Apart from direction, the magnitude of the electrophoretic velocities of the SM particles are also significantly different. Figure \ref{fig:figure4}(c) shows the variation of velocities of all four defect types mentioned above.  It varies quadratically with field, but with different slopes for four different defect configurations. For a given field, quadrupoles of type-I moves with velocity almost one order of magnitude higher than type-II quadrupoles (comparing Fig. \ref{fig:figure4}(c) (ii) and (iii)).  A comparative analysis of different defect types reveals the fact that, for a given field, a wide range of transport velocities could be achieved by tailoring the defects accompanying the particles. Thus, defect polymorphism of the SM particles leads to manifold transport in terms of direction as well as magnitude.

\section{Conclusion} 
To summarise, anisometric snowman shaped dielectric particles with homeotropic anchoring induce point or Saturn ring defects at four different locations. Specifically, a point or ring defect is nucleated either on the small or big lobe of the SM particles. For particles with ring defects encircling either lobes, the quadrupolar symmetry of the director field is broken and they propel along the director with the direction of motion either parallel or antiparallel to the shape-polarity vector ${\bf\hat{s}}$.  Particles with point defect on the small lobe propel almost two times faster than the particles with point defect on the big lobe. Such shape-dependent defect-polymorphism results in reversible electrophoretic transport with varied magnitudes in velocities.
 We also observed a frequency dependent reversal of propulsion direction of dipolar particles at the onset of low frequency electrohydrodynamic instabilities.  In effect, we have at our disposal two tangible mechanisms for reversing the direction of transport of the particles without altering the sign of the dielectric and conductivity anisotropies of the medium.  The defect guided manifold elecrokinetics demonstrated here is in sharp contrast to that are known for spherically symmetric particles. Our findings uncover novel possibilities of controlling electrophoretic transport of asymmetric particles by exploiting defect-polymorphism.
  Evidently, particles with customised shape are promising candidates for achieving  diversity in electrokinetic motility for applications in microsensors, microactuators and microfluidic devices.\\\\\\
 
 \section{Experimental Section}
 \subsection*{Sample and cell preparation}
 We work with asymmetric Polystyrene microparticles, obtained from Magsphere Inc. (USA). The shape of the particle resembles the snowman (SM) with two lobes having different diameters, $D_{1}=3.0$ and $D_{2}=1.6$ ${\upmu}$m (Fig.\ref{fig:figure1} (a,b)). The separation between the centers of the lobes is $L=1.5$ ${\upmu}$m. The separation factor, defined as $2L/(D_{1}+D_{2})$~\cite{liu}, is equal to 0.69 which is less than their dimeric form where $2L/(D_{1}+D_{2})=1$.  The surface of the SM particles are coated with a thin ($50$ nm) layer of SiO\textsubscript{2}, following a method discussed. 
 
 Further, the particles are coated with N, N-dimetyl-N-octadecyl-3 aminopropyltrimethoxysilyl chloride (known as DMOAP) which provides normal or homeotropic  anchoring of LC molecules. A room temperature nematic liquid crystal (MJ 98468, Merck), having birefringence $\Delta \text{n}$ = 0.077 was used in the experiment. In the low frequency regime ($<200$ Hz), it exhibits negative dielectric anisotropy ($\Delta\epsilon$ = -3.4) and positive conductivity anisotropy ($\Delta\sigma\simeq 2\times 10^{-11}$ S/m) (Figure S2, Supporting Information). A small quantity ($\approx$ 0.01 wt$\%$) of DMOAP coated SM particles was dispersed  in the nematic phase through vortex mixing and sonication. The experimental cells were made of two ITO (indium-tin-oxide) coated glass plates. The plates were spin-coated with polyimide AL-1254 (JSR Corporation, Japan) and cured at $180^{\circ}$C, and then rubbed uni-directionally using a bench-top rubbing machine (HO-IAD-BTR-01) to obtain a planar alignment of the nematic director $\bf{\hat{n}}$. Two such plates were assembled, similar to the parallel plate capacitor configuration and fixed with a UV curable adhesive, mixed with spherical glass beads (diameter 8 $\upmu$m). The actual cell gap was measured through an interferometric technique. Electrical contacts to the cells were provided by soldering two thin and flexible wires onto the ITO substrates.
 
  \subsection*{Procedure for Silica coating}
  Initially, 3 gm of PVP was added to 30 mL of 1-pentanol and sonicated well for 2 hours. Then, 200 $\upmu$L of PS particles was added to the above solution an dispersed homogeneously after which, 3 mL of ethanol, 840 $\upmu$L of milliQ water and 200 $\upmu$L of 0.18 M sodium citrate dihydrate were added to the dispersion and shaken thoroughly for 3 minutes. In the next step, 675 $\upmu$L of ammonia solution (25 wt\%) was added to it. After 5 minutes, 500 $\upmu$L of TEOS was added to the mixture and shaken gently for 3 minutes. The reaction was continued for 15 minutes. Immediately after 15 minutes, the particles were washed a few times with ethanol to obtain silica coated snowman particles.
 
 \subsection*{Experimental setup}
\indent We used an inverted polarizing optical microscope (Nikon eclipse Ti-U) with a 60x water immersion objective (Nikon, NIR Apo 60/1.0). A first-order full-wave retardation plate or $\lambda$-plate (530 nm) was employed for constructing the director profile surrounding the SM particles. 
A function generator (AFG 3102, Tektronix) connected to a voltage amplifier (TEGAM 2350) was used for applying ac electric field. A charge-coupled device (CCD) video camera (iDs-UI) attached to the microscope was used to record the particle trajectories at a frame rates of 20 - 100 per second. A particle tracking program was used off-line to track the position of the particles. A Fiji Imagej plugin Trackmate is used for creating colour coded trajectories of particles~\cite{tek}   \\\\\\\\\\\


    
{\noindent \bf Supporting Information}\newline
Supplementary materials are available online or from the author.\\

 {\noindent \bf Acknowledgments}
 \newline
SD gratefully acknowledges the support from the DST (DST/SJF/PSA-02/2014-2015), and University of Hyderabad (UoH/IoE/RC1- 20-010). DVS.  and DKS acknowledges DST for INSPIRE fellowship. R.K.P. acknowledges the Department of Science and Technology for INSPIRE Faculty Award Grant [DST/INSPIRE/04/2016/002370] and a Core Research Grant (CRG/2020/006281, DST-SERB), Government of India, for funding.\\

{\noindent\bf Conflict of Interest}
\newline
The authors declare no conflict of interest.\\

{\noindent \bf Keywords}
\newline
 Microscale, Material science, Physics, Topological defects, Liquid crystals, Electrophoresis, Anisometric colloids, Transport \\


\begin{thebibliography}{99}


\bibitem{morgan} H. Morgan and N. G. Green, \textit{AC Electrokinetics: Colloids and Nanoparticles} (Research Studies Press Ltd, 2003).

\bibitem{ramos} A. Ramos, \textit{Electrokinetics and Electrohydrodynamics in Microsystems} (Spinger, 2011).


\bibitem{drop} S. V. Dorp,  U. F. Keyser, N. H. Dekker, C. Dekker, and S. G. Lemay, Nat. Phys. {\bf 2009}, 5, 347.


\bibitem{alex} A. Terray, J. Oakey, and D. W. M. Marr, Science \textbf{296}, 1841 (2002).

\bibitem{div2} A. F. Demir{\"o}rs, F. Eichenseher, M. J. Loessner,  and A. R. Studart,  Nat. Commun. {\bf 2017}, 8, 1872.


\bibitem{rc} R. C. Hayward, D. A. Saville, and I. A. Aksay, Nature {\bf 2000}, 404, 56.


\bibitem{mur} V.A. Murtsovkin, Colloid Jour. {\bf 1996}, 58, 341. 

\bibitem{tm} T. M. Squires and S. R. Quake, Rev. Mod. Phys. {\bf 2005}, 77, 977.

\bibitem{baz1} M. Z. Bazant, M.S. Kilic, B. D. Storey, and A. Ajdari, Adv. Colloid Interface Sci. {\bf 2009}, 152, 48.

\bibitem{baz2} T. M. Squires and M. Z. Bazant, J. Fluid Mech. {\bf 2004}, 509, 217.

\bibitem{baz3} M.Z. Bazant and T. M. Squires, Phys. Rev. Lett. {\bf 2004}, 92, 066101.

\bibitem{lub} P. Poulin, H. Stark, T. C. Lubensky, and D. A. Weitz, Science {\bf 1997}, 275, 1770.

\bibitem{stark} H. Stark, Phys. Rep. {\bf 2001}, 351, 387.
\bibitem{igor} I. Mu\v{s}evi\v{c}, M. \v{S}karabot, U. Tkalec, M. Ravnik, and S. \v{Z}umer, Science {\bf 2006}, 313, 954.

\bibitem{im} I. Mu\v{s}evi\v{c}, \textit{Liquid crystal colloids}. 2017. (Springer Inter- national Publishing AG).

\bibitem{is} I. I. Smalyukh, Annu. Rev. Condens. Matter Phys. {\bf 2018}, 9, 207.

\bibitem{od} O. D. Lavrentovich, Curr. Opin. Colloid Interface Sci. {\bf 2016}, 21, 97.

\bibitem{od1}O. D. Lavrentovich, Soft Matter {\bf 2014}, 10, 1264.


\bibitem{oleg2} I. Lazo, C. Peng, J. Xiang, S. V. Shiyanovskii, and O. D. Lavrentovich, Nat. Commun. {\bf 2014}, 5, 5033 .

\bibitem{oleg} I. Lazo and O. D. Lavrentovich, Phil. Trans. Soc. A {\bf 2013}, 371, 2012255.

\bibitem{oleg1} O. D. Lavrentovich, I. Lazo, and O. P. Pishnyak, Nature {\bf 2010}, 467, 947.

\bibitem{sd} D. K. Sahu, S. Kole, S. Ramaswamy and S. Dhara, Phys. Rev. Res. {\bf 2020}, 2, 032009(R).

\bibitem{sd1} D. K. Sahu, and S. Dhara, Phys. Rev. Applied. {\bf 2020}, 14, 034004.

\bibitem{sd2} D. K. Sahu, and S. Dhara, Phys. Fluids {\bf 2021}, 33, 0187106.

\bibitem{david1} S. Sacanna, D. J. Pine, Curr. Opin. Colloid Interface Sci. {\bf 2011}, 16, 96.

\bibitem{david2} S. Sacanna, M. Korpics, K. Rodriguez, L. Col{\'o}n-Mel{\'e}ndez, S-H Kim, D. J. Pine and Gi-Ra Yi, Nat. Commun. {\bf 2013}, 4. 1688.

\bibitem{liu} M. Liu, F. Dong, N. S. Jackson, M. D. Ward, and M. Weck, J. Am. Chem. Soc. {\bf 2020}, 142, 16528.

\bibitem{3D}C. Zhu, A. J. Pascall, N. Dudukovic,
M. A. Worsley, J. D. Kuntz, E. B. Duoss and C. M. Spadaccini, Annu. Rev. Chem. Biomol. Eng., {\bf 2019}, 10 17.

\bibitem{science} C. Lapointe, T. G. Mason and I. I. Smalyukh. Science {\bf 2009}, 326, 1083.

\bibitem{jaya} J. Dontabhaktuni, M. Ravnik and S. Zumer, Proc. Natl. Acad. Sci. (USA), {\bf 2014}, \textbf 111, 2464.

\bibitem{sh-1} B. Senyuk, M. C. M. Varney, J. A. Lopez, S. Wang, N. Wu and I. I. Smalyukh, Soft Matter, {\bf 2014}, 10, 6014.

\bibitem{sh-2} D. Andrienko, M. P.  Allen, G. Ska\v{c}ej,  and S. \v{Z}umer,   Phys. Rev. E {\bf 2002}, 65, 041702.

\bibitem{sh-3} C. P. Lapointe,   K. Mayoral, and  T. G. Mason,  Soft Matter {\bf 2013}, 9, 7843.

\bibitem{sh-4} M. V. Rasna,  K. P. Zuhail,  U. V.  Ramudu,  R. Chandrasekar,   J. Dontabhaktuni,  and S. Dhara,  Soft Matter {\bf 2015}, 11, 7674.

\bibitem{sh-5} B. Senyuk,  Q.  Liu,   S. He, R. D. Kamien, R. B.  Kusner, T. C. Lubensky  and  I. I. Smalyukh,   Nature (London) {\bf 2013}, 493, 200.

\bibitem{sh-6}  M. A. Gharbi,  M. Cavallaro, G. Wu,  D. A. Beller, R. D. Kamien,   S. Yang, and  K. J. Stebe,   Liq. Cryst. {\bf 2013}, 40, 1619.

\bibitem{rasi}  Muhammed Rasi M, R. K. Pujala and S. Dhara,  Sci. Rep. {\bf 2019}, 9, 4652.

\bibitem{sag1} S. Hern{\`a}ndez-Navarro, P. Tierno, J. A. Farrera, J. Ign{\'e}s-Mullol, and F. Sagu{\'e}s, Angew. Chem. Int. Ed. {\bf 2014}, 53, 10696.

\bibitem{sag2} A. V. Straube, J. M. Pag{\'e}s, P. Tierno, J. Ign{\'e}s-Mullol, and F. Sagu{\'e}s, Phys. Rev. Res. {\bf 2019}, 1, 022008(R).

\bibitem{sag3}A. V. Straube, J. M. Pag{\'e}s, A. O. Ambriz, P. Tierno, J. Ign{\'e}s-Mullol, and F. Sagu{\'e}s, New. J. Phys. {\bf 2018}, 20, 02075006.

\bibitem{rasna} M. V. Rasna, U. V. Ramudu, R. Chandrasekar and S. Dhara, Phys. Rev. E {\bf 2017}, 95, 012710.



\bibitem{satya}S. Paladugu, C. Conklin, J. Vinals, and O. D. Lavrentovich, Phys. Rev. Appl. {\bf 2017}, 7, 034033.

\bibitem{ning}F. Ma, X. Yang, H. Zhao and N. Wu,  Phys. Rev. Lett. {\bf 2015}, 115, 208302.

\bibitem{pg} P. G. de Gennes, \textit{The Physics of Liquid Crystals} (Clarendon, Oxford,1993).

\bibitem{blinov} L. M. Blinov, Sci. Prog., Oxf. {\bf 1986}, 70, 263.

\bibitem{sasaki1} Y. Sasaki, Y. Takikawa, V. S. R. Jampani, H. Hoshikawa, T. Seto, C. Bahr, S. Herminghaus, Y. Hidakac and H. Orihara. Soft Matter {\bf 2014}, 10,  8813.

\bibitem{sasaki2} Y. Nishioka, F. Kobayashi, N. Sakurai, Y. Sasaki and H. Orihara, Liq. Cryst. {\bf 2016}, 43, 427.

\bibitem{taka}K. Takahashi and Y. Kimura, Phy. Rev. E {\bf 2014}, 90, 012502.
\bibitem{acharya} G. R. Acharya, Ph. D Thesis:\textit{Electroconvection and Pattern Formation in Nematic Liquid Crystal} (2009).

\bibitem{tek}D. Ershov, M. Phan, J. W. Pylv{\"a}n{\"a}inen, S. U. Rigaud, L. Le Blanc, A. Charles-Orszag, J. R. W. Conway, R. F. Laine, N. H. Roy, D. Bonazzi, G. Duménil,  G. Jacquemet, and J. Tinevez. doi: https://doi.org/10.1101/2021.09.03.458852.




\end {thebibliography}
\end{document}